\documentclass[9pt,twocolumn,twoside]{osajnl}

\journal{josaa} 
\setboolean{shortarticle}{false}

\newcommand{\xdir}{$\hat{\mathbf{x}}$ }
\newcommand{\ydir}{$\hat{\mathbf{y}}$ }
\newcommand{\zdir}{$\hat{\mathbf{z}}$ }

\newcommand{\hatx}{\hat{\mathbf{x}}}
\newcommand{\haty}{\hat{\mathbf{y}}}
\newcommand{\hatz}{\hat{\mathbf{z}}}

\newcommand{\rpos}{\mathbf{r}}
\newcommand{\rposp}{\mathbf{r}'}

\newcommand{\kt}{{\mathbf{k}_t}}
\newcommand{\ktp}{\mathbf{k}'_t}
\newcommand{\ktpp}{\mathbf{k}''_t}
\newcommand{\ktmn}{\mathbf{k}_t^{mn}}

\newcommand{\Eincr}{\mathbf{E}_\text{inc}}
\newcommand{\Einc}{\tilde{\mathbf{E}}_\text{inc}}

\newcommand{\EincTE}{\tilde{E}_{\text{inc},TE}}
\newcommand{\EincTM}{\tilde{E}_{\text{inc},TM}}
	
\newcommand{\waveTE}{\mathbf{E}^{TE}_\text{inc}}
\newcommand{\waveTM}{\mathbf{E}^{TM}_\text{inc}}
	
\newcommand{\Eifk}{{\underline{\underline{E}}_\text{tot}^0}}
\newcommand{\Hifk}{{\underline{\underline{H}}_\text{tot}^0}}

\newcommand{\W}{\tilde{\underline{\underline{W}}}}
\newcommand{\Wnotilde}{{\underline{\underline{W}}}}

	\newcommand{\Ptot}{P_\text{tot}}
	\newcommand{\Po}{\underline{\underline{P}}_0}
	\newcommand{\Hmn}{\underline{\underline{\mathcal{H}}}_{mn}}
	
	\newcommand{\Cmn}{\underline{\underline{C}}}
	
\newcommand{\vol}{\Omega}
\newcommand{\voluc}{{\Omega_{0}}}

\newcommand{\eigP}{\Lambda_a}
\newcommand{\vectP}{\bm{\mathcal{E}}_a}

\newcommand{\Smat}{{\underline{\underline{S}}}}

\usepackage[normalem]{ulem}
\usepackage{cancel}
\usepackage{bm}
\usepackage{gensymb} 
\ifthenelse{\boolean{shortarticle}}{\colorlet{color2}{color2b}}{\colorlet{color2}{color2}} 

\title{Identification of the absorption processes in periodic plasmonic structures using Energy Absorption Interferometry}

\author[1*]{Denis Tihon}
\author[2]{Stafford Withington}
\author[2]{Christopher N. Thomas}
\author[1]{Christophe Craeye}

\affil[1]{Universit\'{e} catholique de Louvain, ICTEAM Institute, Place du Levant 3, 1348 Louvain-la-Neuve, Belgium}
\affil[2]{Cambridge University, Cavendish Laboratory, J.J. Thomson Avenue, CB3 OHE Cambridge, UK}

\affil[*]{Corresponding author: denis.tihon@uclouvain.be}

\dates{Compiled \today}

\ociscodes{(000.4430) Numerical approximation and analysis; (010.5620) Radiative transfer; (030.4070) Modes; (050.1755) Computational electromagnetic methods; (300.1030) Absorption.}

\doi{\color{blue}\textcopyright 2019 Optical Society of America. One print or electronic copy may be made for personal use only. Systematic reproduction and distribution, duplication of any material in this paper for a fee or for commercial purposes, or modifications of the content of this paper are prohibited. The final (published) version is available at \url{https://doi.org/10.1364/JOSAA.36.000012}}

\begin{abstract}
Power dissipation in electromagnetic absorbers is a quadratic function of the incident fields. To characterize an absorber, one needs to deal with the coupling that may occur between different excitations. Energy Absorption Interferometry (EAI) is a technique that highlights the independent degrees of freedom through which a structure can absorb energy: the natural absorption modes of the structure. The coupling between these modes vanishes. 
In this paper, we use the EAI formalism to analyse different kinds of plasmonic periodic absorbers while rigorously accounting for the coupling: resonant golden patches on a grounded dielectric slab, parallel free-standing silver wires and a silver slab of finite thickness. The EAI formalism is used to identify the physical processes that mediate absorption in the near and far field. First, we demonstrate that the angular absorption, which is classically used to characterize periodic absorbers in the far field and which neglects the coupling between different plane waves, is only valid under stringent conditions (subwavelength periodicity, far field excitation and negligible coupling between the two possible polarizations). Using EAI, we show how the dominant absorption channels can be identified through the signature of the absorption modes of the structure, while rigorously accounting for the coupling. We then exploit these channels to improve absorption. We show that long-range processes can be exploited to enhance the spatial selectivity, while short-range processes can be exploited to improve absorptivity over wide angles of incidence. Lastly, we show that simply adding scatterers with the proper periodicity on top of the absorber, the absorption can be increased by more than one order of magnitude.
\end{abstract}

\setboolean{displaycopyright}{true}

\begin{document}

\maketitle
\thispagestyle{fancy}

\ifthenelse{\boolean{shortarticle}}{\ifthenelse{\boolean{singlecolumn}}{\abscontentformatted}{\abscontent}}{}

\section{Introduction}
There are many applications in which the absorption of electromagnetic fields must be carefully controlled. For example, in photovoltaic technology better absorption of the fields leads to a better efficiency \cite{Atwater2010}. We can also cite medical applications, where the specific absorption rate must be precisely controlled in order either to limit it and avoid a temperature rise inside the body of the patient (see e.g. \cite{deGreef2013}), or to the contrary, maximize it at a given location within the body (e.g. a tumor) in order to weaken or damage this region \cite{Rao2010, Iero2017, Nguyen2017}. Lastly, imaging systems working in the infrared and visible spectrum are generally based on bolometers, so that a precise control of the absorbing capabilities of the detectors (selectivity and amplitude of the absorption) can lead to important improvements in the image resolution. 

For all the applications of this non-exhaustive list, a complete and efficient characterization of the absorption capabilities of a given structure is essential.
Characterizing the way in which a structure can scatter incident fields is facilitated by the fact that Maxwell's equations are linear with respect to the source terms. However, when it comes to absorption, the power dissipated is a quadratic quantity of the fields, such that superposition based on linearity can no longer be considered by default. The characterization of absorption may therefore become more complicated and physical insight into the phenomena through which absorption takes place may be partially or totally lost. 

In many papers (see e.g. \cite{Liu2010, Hao2011, Alu2011}), the quality of a periodic absorber is assessed by showing its plane wave absorption versus angle of incidence. However, as soon as sources are located in the near field or the periodicity of the absorber becomes too large, this information is not sufficient to fully describe the absorption capabilities of the structure because it neglects the coupling between different plane waves. Moreover, even for far-field excitations and a periodicity smaller than half the wavelength, the absorber may be sensitive to a given polarization of the fields (e.g. circular \cite{Ye2016} vs. linear \cite{Xiong2013}) which cannot be known \textit{a-priori}.

In previous papers, some of the authors developed a formalism to treat the absorption of partially coherent fields \cite{Saklatvala2006}. 
Based on this formalism, they developed Energy Absorption Interferometry (EAI), an experimental technique that allows to fully and efficiently characterize the way a structure can absorb energy \cite{Withington2007, experimental_setup}. The technique consists in illuminating the sample using two sources whose positions, orientations and relative phase can be controlled. For fixed positions and orientations, the total power absorbed by the sample exhibits fringes when the relative phase is varied, highlighting the correlation between the sources. By measuring and storing the correlation function for various positions and orientations, one can find all the information required to fully characterize the absorption capabilities of the structure.

One of the key points of this formalism is the fact that lossy structures are absorbing energy through independent and orthogonal absorption modes. Any structure can be completely characterized through the spatial form and amplitude of its natural absorption modes, leading to a compact and intuitive formulation. 
Among the benefits of the proposed formalism are an easy treatment of partially spatially coherent incident fields, such as those originating from thermal sources, or the possibility to retrieve the absorption modes of a structure from experiments \cite{experimental_setup}, provided that the power dissipated by the structure can be measured e.g. through an increase of its temperature. 

In \cite{Craeye2014, Tihon2016}, EAI was extended to the study of periodic structures with near-field and far-field excitations. It was shown that, imposing the quasi-periodicity of the fields (i.e. periodicity of the fields within a linear phase-shift), the study of the structure can be constrained to a single unit cell (Bloch modes). The Bloch modes characterized by different phase shifts are orthogonal to each other. 

In this paper, we use the EAI formalism to provide a new perspective on the physical processes that mediate absorption in plasmonic structures through three examples: a perfect infrared absorber based on resonant gold patches \cite{Liu2010}, a silver slab and a periodic structure made of parallel free-standing resonant silver rods \cite{Ono2005}. The first example is used to demonstrate that the characterization of an absorber through its plane wave absorption is not sufficient because it neglects the coupling that may occur between plane waves. It is shown how using the EAI formalism solves the problem by rigorously accounting for that coupling. Then, the two other examples are used to illustrate the information that can be gathered through a full (real or numerical) EAI experiment. It is also shown how this information can be used to identify the processes mediating the absorption at subwavelength scales and how to harness these to control absorption. Some preliminary results were presented in \cite{Tihon2017}. 

In Section 2, for self-consistency, the formalism used is briefly summarized.
Then, in Section 3, we study a modified version of the absorber of \cite{Liu2010}. The structure illustrates the need for EAI in order to completely characterize the absorption capabilities of the structure. 
In Section 4, we show how a deep physical insight into the near field processes that mediate absorption in the two last absorbers can be gained from the quantities measured by the EAI. It is shown that in the very near field, the wire medium is mainly absorbing fields through localized surface modes while the absorption within the silver slab is mainly due to the long-range excitation of surface plasmons (SP). Section 5 serves as a proof of concept: exploiting this knowledge, the absorption of the wire medium is increased by more than one order of magnitude by simply adding a grating with the proper period on top of the wires.

\section{Dissipation in lossy structures}
\label{sec:theory}
In this section we rapidly remind the formalism used to obtain the natural absorbing modes of a structure periodic along two directions. For a more detailed treatment, please refer to \cite{Withington2007, Craeye2014, Tihon2016}.

\subsection{General framework}
\label{sec:GF}
We consider a structure illuminated by quasi-monochromatic fields, with angular frequency $\omega$. A phasor notation will be used hereafter with an implied $\exp(j \omega t)$ time dependency of the fields.

The structure is periodic along two orthogonal directions, denoted \xdir and \ydir, with periodicities $a$ and $b$ respectively, and semi-infinite along along the \zdir direction. The structure is made of isotropic material and is characterized by its complex periodic permittivity $\varepsilon(\rpos)$ and permeability $\mu(\rpos)$, whose imaginary parts account for all the losses mechanisms (dielectric loss, conduction loss, magnetic loss). To these permittivity and permeability one can associate real electric and magnetic equivalent conductivities $\sigma_J = -\omega \mathcal{I}\{\varepsilon\}$ and $\sigma_M = -\omega \mathcal{I}\{\mu\}$, respectively, with $\mathcal{I}\{a\}$ the imaginary part of $a$. The structure is located in the half-space $z<0$ and for the sake of simplicity, we will only consider fields due to sources located in the half-space $z>h$, with $h$ a positive constant. In order to express the incident fields, a plane wave decomposition is used hereafter. However, as explained in Section \ref{sec:diff_exc}, the formalism presented can be reformulated using any basis for the field. Using the plane wave decomposition, the incident fields in the half-space $z<h$ ($\Eincr(\rpos)$) read: 
\begin{equation}
\label{eq:28-09-18-01}
\begin{split}
\Eincr(\rpos) = \dfrac{1}{4\pi^2} \iint \Big(&\EincTE(\kt) \waveTE(\rpos|\kt) \\
&+
\EincTM(\kt) \waveTM(\rpos|\kt) \Big) d\kt,
\end{split}
\end{equation}
with $\EincTE(\kt)$ and $\EincTM(\kt)$ the amplitudes of the incident TE and TM plane waves of real transverse wavevector $\kt$ and
\begin{subequations}
\begin{align}
\waveTE(\rpos|\kt) &= \dfrac{1}{k_t} (-k_y \hatx + k_x \haty) \exp^{-j(k_x x + k_y y - \gamma (z-h))}\\
\waveTM(\rpos|\kt) &= \dfrac{\gamma}{k_t k_0} (k_x \hatx + k_y \haty + k_t^2/\gamma ~ \hatz) \exp^{-j(k_x x + k_y y - \gamma (z-h))}
\end{align}
\end{subequations}
with $k_0 = \omega \sqrt{\varepsilon_0 \mu_0}$ the free-space wave vector, $\kt = (k_x, k_y, 0)$, $k_t = |\kt|$ the amplitude of $\kt$ and 
\begin{equation}
\label{eq:10-07-01}
\left\{
\begin{split}
\gamma &= \sqrt{k_0^2-k_t^2}~~~~  &\text{if } k_0>k_t \\
\gamma &= -j \sqrt{k_t^2-k_0^2}~~~~  &\text{otherwise}.
\end{split}
\right.
\end{equation}
Note that, in this paper, integrations whose domain of integration is not stated explicitly are carried out over the whole space.

 Using the spectral decomposition of the incident fields in \eqref{eq:28-09-18-01} and the superposition principle, the total fields can be represented as:
\begin{subequations}
\label{eq:13-04-01}
\begin{align}
\mathbf{E}(\rpos) &= \dfrac{1}{4\pi^2} \iint \Eifk(\rpos|\kt) \cdot \Einc(\kt) d\kt \\
\mathbf{H}(\rpos) &= \dfrac{1}{4\pi^2} \iint \Hifk(\rpos|\kt) \cdot \Einc(\kt) d\kt
\end{align}
\end{subequations}
with $\Einc = (\EincTE, \EincTM)$ and $\Eifk(\rpos|\kt)$ and $\Hifk(\rpos|\kt)$ the tensor describing the total (incident + scattered) electric and magnetic fields induced at point $\rpos$ by incident plane waves with transverse wave vector $\kt$. 

The absorbed power density at a given position in the material, $P(\rpos)$, only depends on the local value of the fields and is given by
\begin{equation}
\label{eq:13-04-02}
P(\rpos) = \dfrac{1}{2} \Big(\sigma_J(\rpos) \mathbf{E}^*(\rpos) \cdot \mathbf{E}(\rpos) + \sigma_M(\rpos) \mathbf{H}^*(\rpos) \cdot \mathbf{H}(\rpos) \Big)
\end{equation}
with $A^*$ the complex conjugate of $A$, $\mathbf{E}$ and $\mathbf{H}$ the total electric and magnetic fields, respectively. 

Substituting (\ref{eq:13-04-01}) into (\ref{eq:13-04-02}) and integrating over a chosen volume $\vol$, the total power dissipated within this volume is given by \cite{Tihon2016}
\begin{equation}
\label{eq:14-04-01}
\Ptot = \dfrac{1}{16\pi^4} \iint \iint \W(\kt, \ktp) : \Po(\kt, \ktp) d\kt d\ktp
\end{equation}
with $\Po$ and $\W$ the cross-spectral power density and the cross-spectral field density functions, respectively, that read
\begin{equation}
\label{eq:W}
\W(\kt, \ktp) = \Big< \Einc^*(\kt) \Einc(\ktp) \Big>,
\end{equation}
\begin{equation}
\begin{split}
\label{eq:P0}
\Po(\kt, \ktp) = \dfrac{1}{2} \iiint_\vol \Big(&\sigma_J \Eifk^\dagger(\rpos|\kt) \cdot \Eifk(\rpos|\ktp) \\
&+ \sigma_M \Hifk^\dagger(\rpos|\kt) \cdot \Hifk(\rpos|\ktp) \Big) dV,
\end{split}
\end{equation}
with $<a>$ the ensemble average of $a$, $\underline{\underline{A}}^\dagger$ the transpose conjugate of $\underline{\underline{A}}$, $\underline{\underline{A}} = (\mathbf{a} \mathbf{b})$ the dyadic product of $\mathbf{a}$ and $\mathbf{b}$ ($A_{ij} = a_ib_j$) and $\underline{\underline{A}}: \underline{\underline{B}} = \text{Trace}\{\underline{\underline{A}}^T \cdot \underline{\underline{B}} \}$ the double contraction of dyads $\underline{\underline{A}}$ and $\underline{\underline{B}}$, with $\underline{\underline{A}}^T$ the transpose of $\underline{\underline{A}}$. In (\ref{eq:W}), the ensemble average of the fields is used to handle the stochastic nature of the incident fields, when they are only partially coherent in space \cite{Wolf2003, Tihon2016}. Equation (\ref{eq:P0}) is the extension of Equation (10) of \cite{Tihon2016} to magnetic losses.

If the region $\vol$ over which the absorption is computed shares the same periodicity as the lattice, it can be shown that the study of the structure can be limited to only one unit-cell with quasi-periodic boundary conditions (i.e. periodic conditions within a linear phase-shift between consecutive unit-cells) \cite{Craeye2014}. Denoting $\voluc$ the restriction of $\vol$ to one unit cell and performing a change of coordinates in the spectral domain, the cross-spectral power density function (\ref{eq:P0}) becomes \cite{Tihon2016}
\begin{equation}
\label{eq:14-04-03}
\Po(\kt-\ktpp, \kt+\ktpp) = \sum_{m,n = -\infty}^{\infty} \Hmn(\kt) \delta(\ktpp-\ktmn)
\end{equation}
with
\begin{equation} 
\ktmn = \Big(m\dfrac{\pi}{a}, n\dfrac{\pi}{b}, 0 \Big)
\label{eq:19-07-01}
\end{equation}
and
\begin{equation}
\begin{split}
\label{eq:Hmn}
\Hmn(\kt) = &\dfrac{1}{2} \iiint_\voluc \Big(\sigma_J \Eifk^\dagger(\rpos|\kt-\ktmn) \cdot \Eifk(\rpos|\kt+\ktmn) \\
&+ \sigma_M \Hifk^\dagger(\rpos|\kt-\ktmn) \cdot \Hifk(\rpos|\kt+\ktmn) \Big) dV.
\end{split}
\end{equation}

The advantages of this specialized formulation are both that the study of the infinite material is limited to one unit-cell, and that the 4D function $\Po$ is reduced to a 2D discrete set of 2D functions $\Hmn$. Moreover, in practice, the terms of the infinite sum of (\ref{eq:14-04-03}) decrease exponentially for $h>0$, so that the sum can be truncated \cite{Craeye2014, Tihon2016}.

\subsection{Natural absorption modes}
\label{sec:NAM}
Using (\ref{eq:P0}), it can be seen that $\Po$ is Hermitian, so that its eigenvalues $\eigP$ are real and its eigenvectors $\vectP$ form an orthogonal basis in which any incident field can be formulated:
\begin{equation}
\label{eq:14-04-04}
\Einc(\kt) = \sum_a \Big( \iint \vectP^*(\ktp) \cdot \Einc(\ktp) d\ktp \Big) \vectP(\kt).
\end{equation}
Combining (\ref{eq:14-04-04}) with the eigenmode decomposition of $\Po$, which reads
\begin{equation}
\Po(\kt, \ktp) = \sum_a \eigP \Big(\vectP^*(\kt) \vectP(\ktp) \Big),
\end{equation}
and using the orthogonality of the eigenmodes, one obtains that the total power absorbed by the structure \cite{Tihon2016}:
\begin{equation}
\label{eq:14-04-05}
\Ptot = \dfrac{1}{16\pi^4} \sum_a \eigP \Big| \iint \vectP^*(\kt) \cdot \Einc(\kt) d\kt \Big|^2.
\end{equation}

The decomposition (\ref{eq:14-04-05}) is very useful from both the physical and practical points of view. From a physical perspective, it means that among the infinite possibilities of incident fields, only the modes whose eigenvalue $\eigP$ is non-negligible contribute to the absorption. Moreover, each mode contributes independently, so that their is no coupling between the different modes. Therefore, using the absorption modes of the structure as a basis to express the incident fields, the power absorption becomes linear with respect to the intensity of the modes. This property may have important applications in problems where the total power dissipated in a given zone should be minimized or maximized (see e.g. \cite{deGreef2013, Rao2010, Iero2017, Nguyen2017}).

From a practical perspective, to fully characterize the absorption capabilities of a structure, one does not need to deal with the full cross-power density function $\Po$, whose interpretation is not intuitive and whose manipulation can be intensive. Decomposing the $\Po$ function into a discrete set of 2D functions $\vectP$ whose associated value $\eigP$ is not negligible offers a compact and intuitive way to present and manipulate the results of the EAI experiment. The incident fields can be directly projected onto these modes in order to obtain the total power absorbed.

We would like to emphasize that the latter fact is remarkable with respect to classical mode-matching techniques in the sense that the absorption modes form an orthonormal basis for the \emph{incident} fields, and not the \emph{total} fields (incident and scattered). For that reason, the computation of the exact amplitude at which one mode is excited does not require any knowledge about the other modes that may exist in the structure, contrarily to classical mode-matching techniques (see e.g. \cite{Collin1990} or \cite{Engheta1990}).

In the case of periodic structures, the set of natural absorption modes of the structure is a function of the phase-shift between consecutive unit-cells.

\subsection{Generalization to arbitrary sources}
\label{sec:diff_exc}
In Subsections \ref{sec:GF} and \ref{sec:NAM}, EAI has been developed for plane-wave excitation. However, any kind of sources can be used to probe the response of the structure. For example, in the experimental setup of \cite{experimental_setup}, dipoles oriented along the \xdir and \ydir directions are used as sources. Instead of considering plane waves across a continuous spectrum $\kt$ with TE and TM polarizations, a continuous range of positions $\rpos$ of \xdir and \ydir oriented dipoles is used to characterize the structure. It is shown in \cite{Craeye2014, Tihon2016} that if the spectrum of the fields emitted by the sources $\Smat(\kt | \rpos)$ as a function of the position of the source and its orientation is known, Equation (\ref{eq:14-04-01}) becomes:
\begin{equation}
\label{eq:03-01-01}
\Ptot = \iint \iint \Wnotilde(\rpos, \rposp) : \Cmn(\rpos, \rposp) d\rpos d\rposp
\end{equation}
with $\Wnotilde$ the correlation function of the source distribution and $C_{ij}(\rpos, \rposp)$ the coupling between a dipole located in $\rpos$ and oriented along the $i$ direction and a dipole located in $\rposp$ and oriented along the $j$ direction. Its formula is given by
\begin{equation}
\Cmn(\rpos,\rposp) = \dfrac{1}{16\pi^4} \iint \iint \Smat(\kt|\rpos)^\dagger \cdot \Po(\kt, \ktp) \cdot \Smat(\ktp|\rposp) d\kt d\ktp.
\end{equation}

The $\Cmn$ function provides relevant information about the typical coherence length within the structure. If the coupling between two sources is non-negligible (i.e. the amplitude of $\Cmn$ is not negligible), it means that the power absorbed by the structure will strongly depend on the relative phase of the sources. To the opposite, if the correlation between the two sources is negligible ($\Cmn(\rpos, \rposp) \rightarrow \underline{\underline{0}}$), it means that the powers absorbed from each source will add incoherently and will therefore be insensitive to their relative phase. 

\section{Conditions of validity of angular absorption spectrum}
Often in the literature, the efficiency of periodic absorbing metasurfaces is assessed by showing the reflection coefficient of incident plane waves on the surface (see e.g. \cite{Liu2010,Hao2011, Alu2011}). This description proves to be sufficient for far field excitation of the metasurface, since the periodicity of the latter is smaller than half a wavelength and the coupling between the two polarizations considered (circular \cite{Ye2016} or linear \cite{Xiong2013}) is negligible. However, that kind of approach is no longer sufficient to provide accurate results as soon as one of the three hypotheses (far field, subwavelength periodicity and negligible cross-polarization) is broken, as shown in this section.

This phenomenon can be explained starting from Equation (\ref{eq:14-04-03}).  Since the structure is periodic, (\ref{eq:14-04-03}) tells us that the cross-spectral power density function $\Po$ is sparse, and that coupling between two plane waves is possible only if the transverse wave vectors of the two plane waves are identical within a vector of the reciprocal lattice of the structure \cite{Craeye2014, Tihon2016}:
\begin{equation}
\label{eq:10-07-02}
 2\ktmn = \kt_{,1} - \kt_{,2}, ~~~~~~ m,n \in \mathbb{Z}
\end{equation}
with $\kt_{,i}$ the transverse wave vector of plane wave $i$. As a reminder, we have previously assumed the fields are emitted from a position $z>h$. Making the hypothesis that $\Hmn$ is bounded for $h > 0$ (which seems reasonable considering the compactness of the radiation operator \cite{Bucci1997} and the presence of losses), (\ref{eq:Hmn}) becomes
\begin{equation}
\label{eq:10-07-03}
\Hmn(\kt|h=h_0) = \Hmn(\kt|h=0^+) \exp\big(-j (\gamma_1+\gamma_2) h_0 \big)
\end{equation}
with $\gamma_i$ as defined in (\ref{eq:10-07-01}), $i=1,2$ corresponding to the transverse wave vectors $\kt-\ktmn$ and $\kt+\ktmn$, respectively. Substituting (\ref{eq:10-07-02}) into (\ref{eq:10-07-03}), considering that 
\begin{equation}
\mathcal{I}(\gamma) \leq k_0-k_t
\end{equation}
and using the Cauchy-Schwartz inequality ($|\kt + 2 \ktmn| \geq 2 k_t^{mn} - k_t$), the following relation can be written:
\begin{equation}
\label{eq:20-07-01}
|\Hmn(\kt|h=h_0)| \leq |\Hmn(\kt|h=0^+)| \exp\big(2(k_0-k_t^{mn}) h_0 \big)
\end{equation}
with $k_t^{mn} = |\ktmn|$. The latter result is important since, for a periodicity smaller than half the wavelength ($a,b < \lambda/2$), the term $k_0-k_t^{mn} <0$ for any pair $(m,n) \neq (0,0)$. It means that, as soon as the source is in the far-field ($h_0 \rightarrow \infty$), the coupling between the incident plane waves vanishes and the $\Po$ operator becomes diagonal. In that case, the absorption modes of the structure, which correspond to the eigenvectors of $\Po$, become the different plane waves themselves and the absorption capabilities of the structure are fully described by providing the way in which each plane wave can be absorbed independently from the others (provided that the coupling between the two polarizations is negligible).
However, if the periodicity of the structure is larger than $\lambda/2$ or if the absorption capabilities of the structure are studied for near-field excitation, the $\Po$ operator is no longer diagonal and the full formalism described in Section \ref{sec:theory} must be used to characterize the absorbing structure.

\begin{figure}[h!]
\center
\includegraphics[width = 7cm]{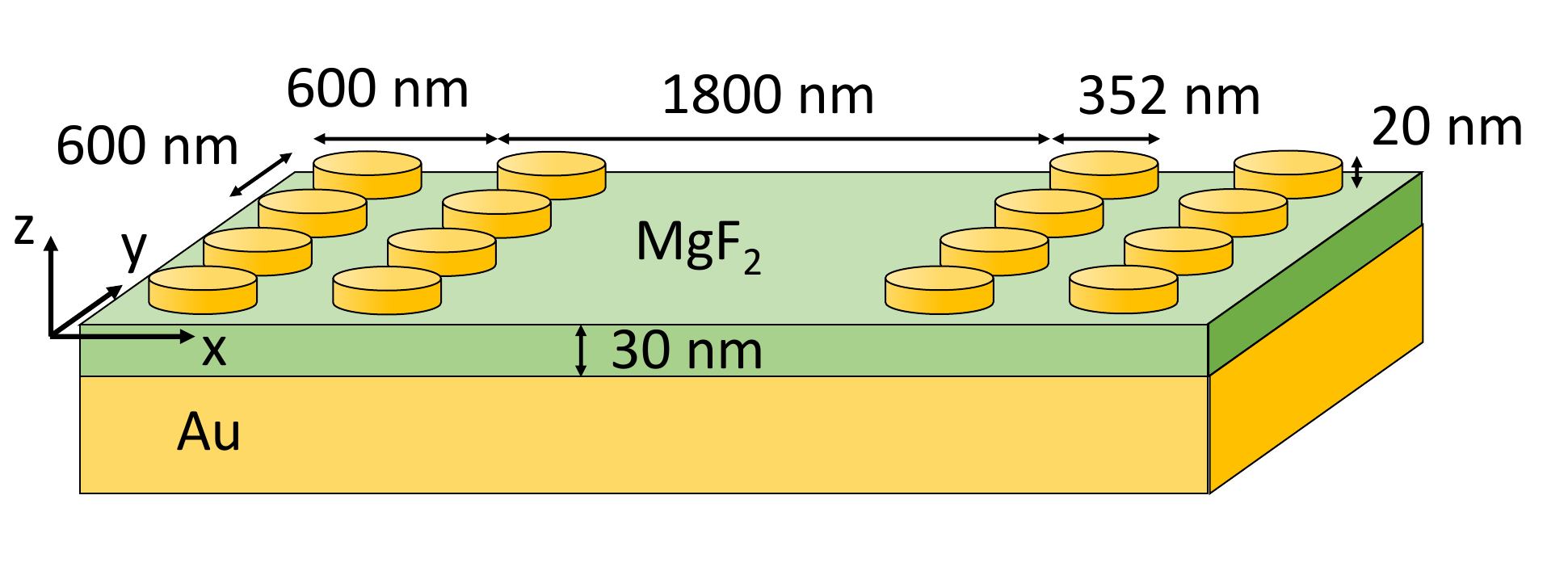}
\caption{Modified version of the absorber of \cite{Liu2010}. It is made of a periodic arrangement of gold patches deposited on a MgF$_2$ spacer and backed with a gold ground plane. The periods along the $\hatx$ and $\haty$ directions are 2400 and 600 nm respectively.}
\label{fig:perfect_absorber}
\end{figure}

In order to illustrate this phenomenon, we considered the structure illustrated in Fig. \ref{fig:perfect_absorber}, which consists of a slightly modified version of the absorber described in \cite{Liu2010}. The original structure is made of gold patches whose periodicities are 600 nm in the $\hatx$ and $\haty$ directions. These patches are deposited on a dielectric spacer made of MgF$_2$ whose relative permittivity is 1.9. The whole layer is backed by an infinite ground plane made of gold. To compute the relative permittivity of the gold, we used the same model as in the original paper, i.e. a Drude model with parameters that have been modified in order to take into account the poor quality of the deposited gold layers \cite{Liu2010}. The free space wavelength of the excitation is chosen to be 1550 nm, which gives a relative permittivity of $-124.6 -12.64j$. In the original paper \cite{Liu2010}, this absorber exhibits a nearly perfect absorption of propagating incident plane waves over a wide range of angles of incidence. Absorption is localized in the resonant cavities formed under each disc.

In order to decrease the absorption to approximately 50\%, we removed two patches every four unit-cells in the $\hatx$ direction. The structure studied can be seen in Fig. \ref{fig:perfect_absorber}. The periodicity in the $\hatx$ direction of the modified structure is greater than half a wavelength. For any fixed phase shift between consecutive unit cells, at least 6 different Floquet modes can propagate in the air (three TE and three TM plane waves), leading to a possible coupling in the absorption between them. 

\begin{figure}[h!]
\center
\begin{tabular}{cc}
\includegraphics[width = 4cm]{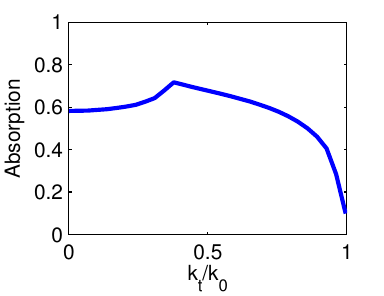} &
\includegraphics[width = 4cm]{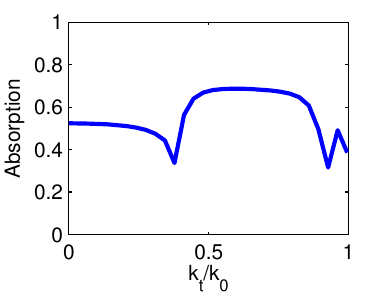} \\
(a) & (b) \\
\includegraphics[width = 4cm]{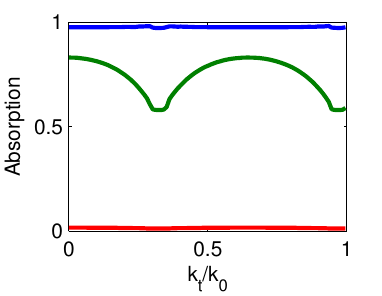} &
\includegraphics[width = 4cm]{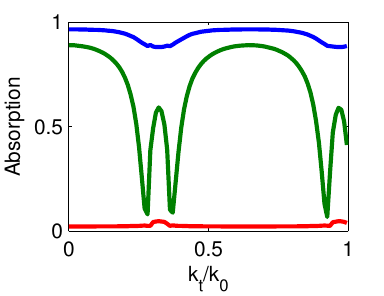} \\
(c) & (d)
\end{tabular}
\caption{Far-field characterization of the structure illustrated in Fig. \ref{fig:perfect_absorber}. (a),(b): Plane wave absorption of the structure. (c),(d): First three natural absorption modes of the structure for different phase-shifts between consecutive unit cells. Graphs (a) and (c) correspond to TE waves while (b) and (d) corresponds to TM waves. }
\label{fig:ZZ_abs}
\end{figure}

The normalized absorption of an incident plane wave as a function of its transverse wave vector $\kt = (0, k_x)$ has been plotted in Figs. \ref{fig:ZZ_abs}(a) and \ref{fig:ZZ_abs}(b) for TE and TM polarized plane waves, respectively. For most of the angles of incidence, the structure absorbs between 40\% and 70\% of the incident power. In Figs. \ref{fig:ZZ_abs}(c) and \ref{fig:ZZ_abs}(d), the first six absorption modes associated with the incident plane waves are shown. Due to the symmetries of the structure and the excitation, it can be shown that the coupling between TE and TM modes vanishes. For ease of interpretation, TE and TM modes have been drawn in separate figures (plot (c) and (d), respectively). Plots (c) and (d) clearly indicate that using a proper combination of propagative incident plane waves, the absorbed power can rise to more than 90\% or drop below 5\%.

Let us illustrate this with an example. Imagine that the structure is illuminated by a plane wave whose intensity is $I_1$. This wave excites several absorption modes of the structure, so that approximately half of the incident power is absorbed. 
Now, using a precise combination of plane waves, it is possible to selectively excite one absorption mode or another. It means that, by properly adjusting the phase and amplitude of the incident plane waves the structure can act as a mirror or a good absorber. Keeping constant $I_1$ and illuminating the structure with additional plane waves, it is possible to decrease the power absorbed by the structure while increasing the incident power, which may seem counter-intuitive. Indeed, using the proper combination of plane waves, the fields can be focused into the places where there is no gold patch, leading to a low absorption.

In this case, since the unit cell is bigger than half a wavelength, the EAI formalism is required to fully characterize the absorbing structure, even for far-field excitation. Note that even for unit cells smaller than half a wavelength, the EAI formalism may be required if the symmetries of the structure and/or the excitation is broken, so that cross-polarization coupling may occur. 

\section{Identification of the absorption processes in silver periodic absorbers}
\begin{figure}[h!]
\center
\begin{tabular}{cc}
\includegraphics[width = 3.5cm]{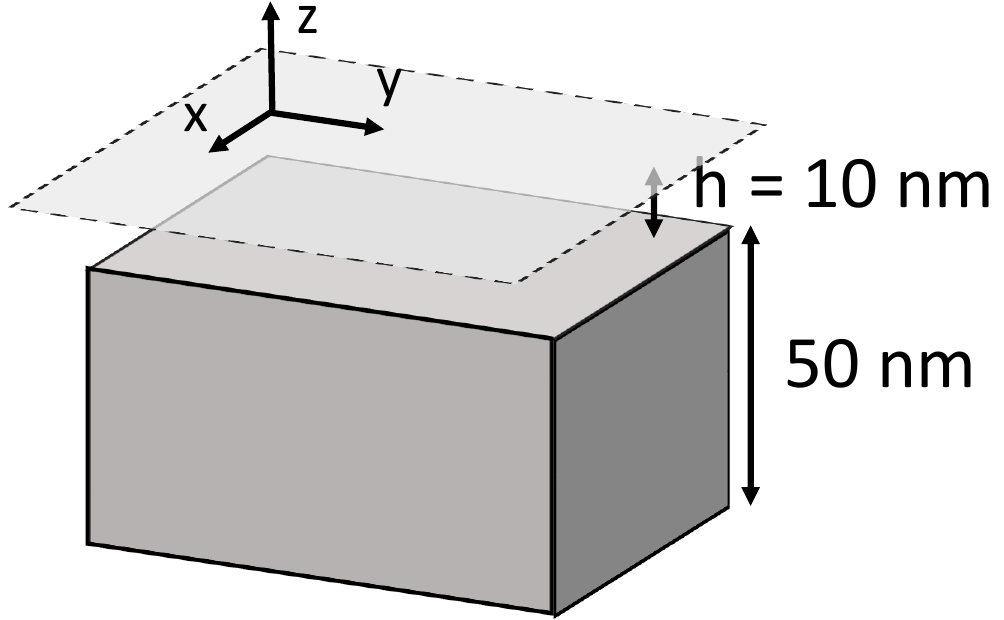}
&
\includegraphics[width = 4cm]{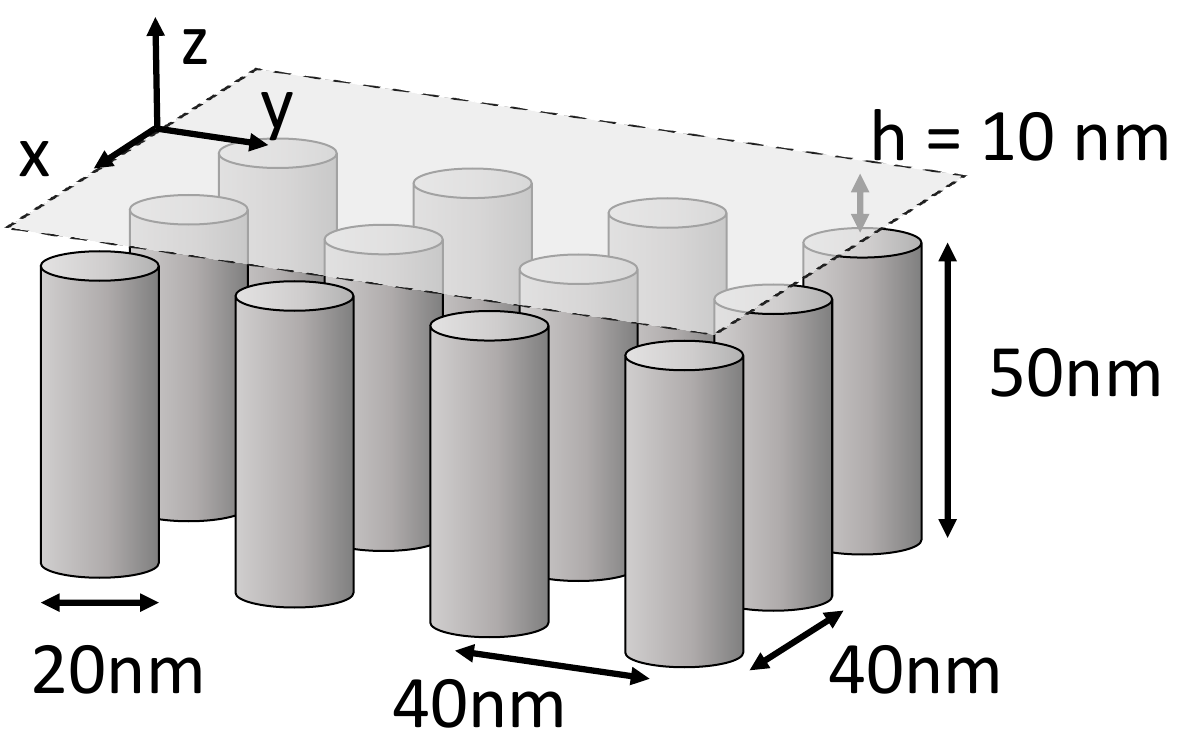}
\\
(a)
&
(b)
\end{tabular}
\caption{The two absorbers studied for near-field sources ($h=10$ nm). (a) The silver slab. (b) The wire medium described in \citep{Ono2005}. The gray plane is the reference plane from which the exciting field is emitted (plane wave or dipole).}
\label{fig:absorbers}
\end{figure}

In order to identify the physical processes that can be exploited to improve absorption, we studied two different kinds of plasmonic absorbers using EAI for an excitation whose free space wavelength is 488 nm. The absorbers have been modeled using a full-wave simulation software based on the Method of Moments \cite{Tihon_simu}. The first one consists of a silver slab of finite thickness. At the frequency studied, silver-air interface is known to support plasmonic surface waves. The wave vector associated with that surface wave is slightly evanescent in free space \cite{SPP} and can propagate over long distances due to the relatively good conductivity of silver at this frequency. The second absorber corresponds to the structure described in \cite{Ono2005}, but arranged in a square lattice (the original structure is hexagonal). It is made of parallel free-standing silver rods. As explained in \cite{Ono2005}, these rods exhibit very localized resonance when they are excited in the near-field. For that reason, they were primarily used for imaging purposes. However, since the response of the structure is very local, each wire is expected to support its own independent absorption mode, so that the number of modes available per unit area is expected to be very high \cite{Narimanov2010}. Moreover, these absorption modes are expected to be localized, since they are based on the individual resonance of the rods rather than collective oscillations within the array. The geometries and their dimensions are illustrated in Figure \ref{fig:absorbers}.

Both structures are characterized for sources located 10 nm away and whose free-space wavelength is 488 nm. The frequency has been chosen to match that of \citep{Ono2005} so that the wire medium exhibits a resonant behaviour.  At this frequency, the relative permittivity of silver is $\varepsilon_r = -9.12-0.304j$ \cite{Johnson1972}.

\begin{figure*}
\center
\begin{tabular}{ccc}
\vspace{-0.2cm}
\includegraphics[width=8.5cm]{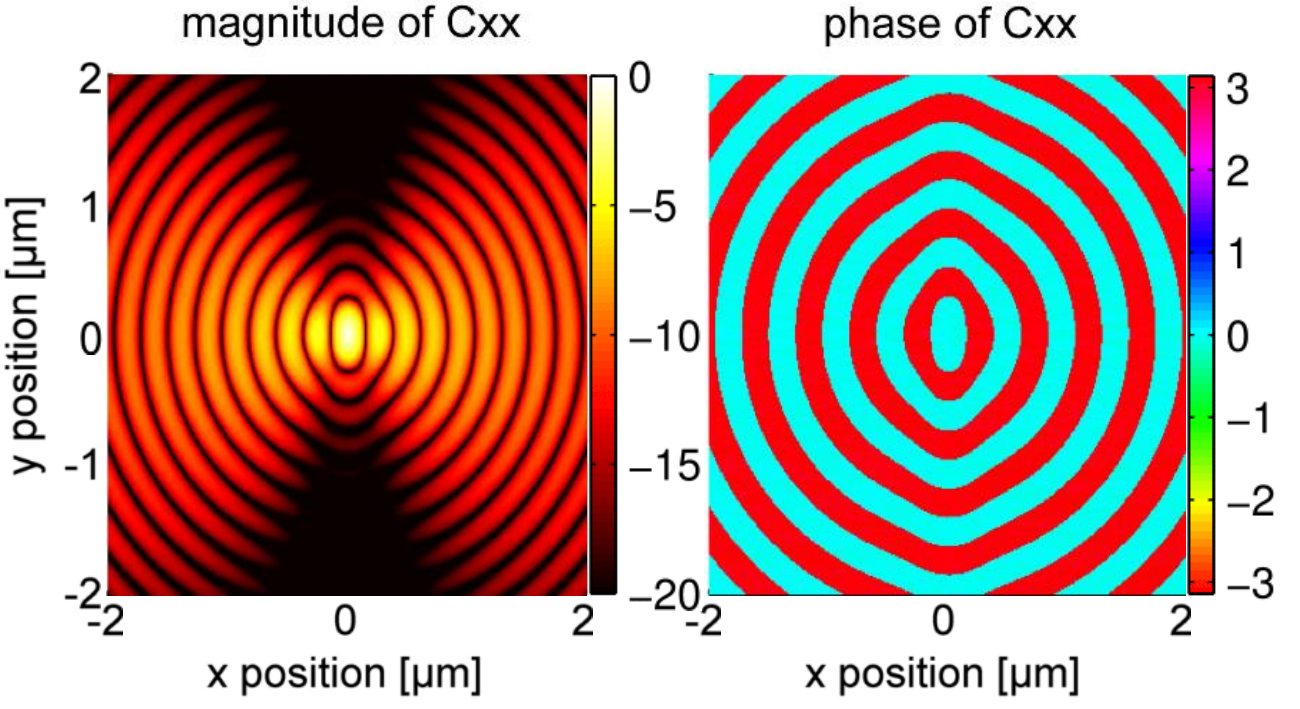} &   &
\includegraphics[width=8.5cm]{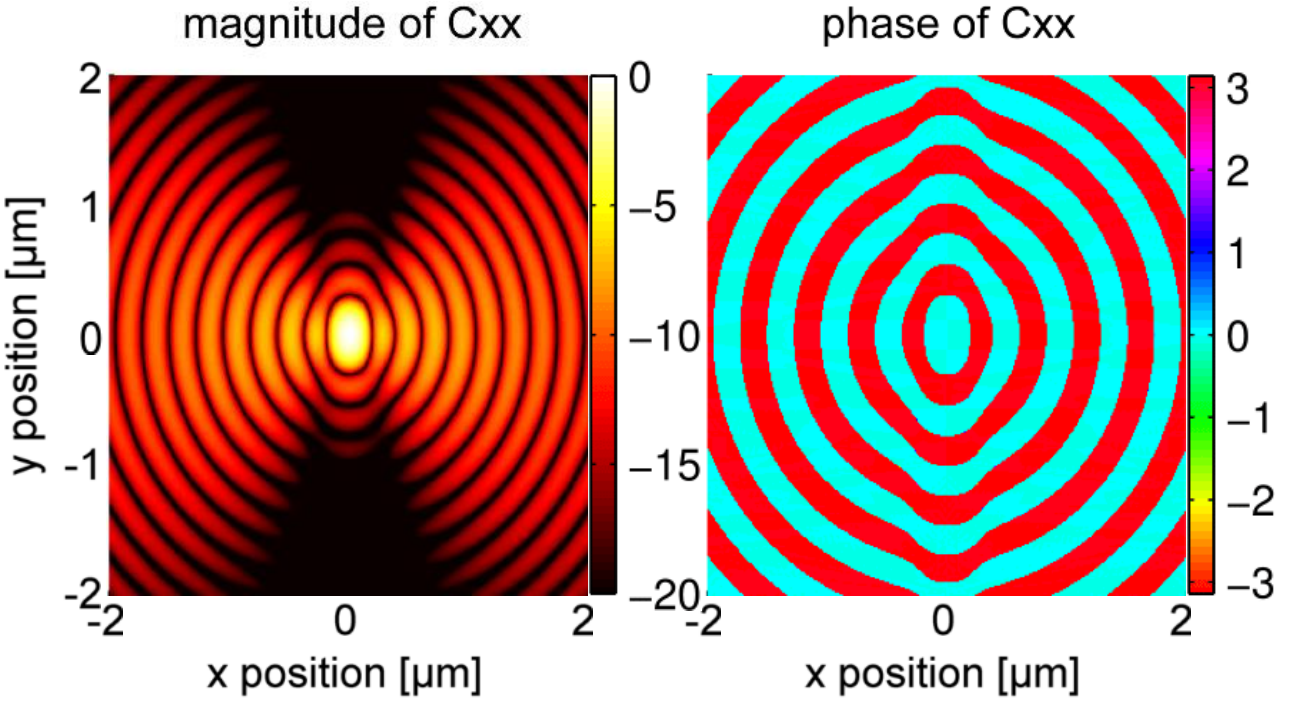} 
\\
\vspace{0.2cm} ~~~~ (a) &  & (b) \\
\vspace{-0.2cm}
\includegraphics[width=8.5cm]{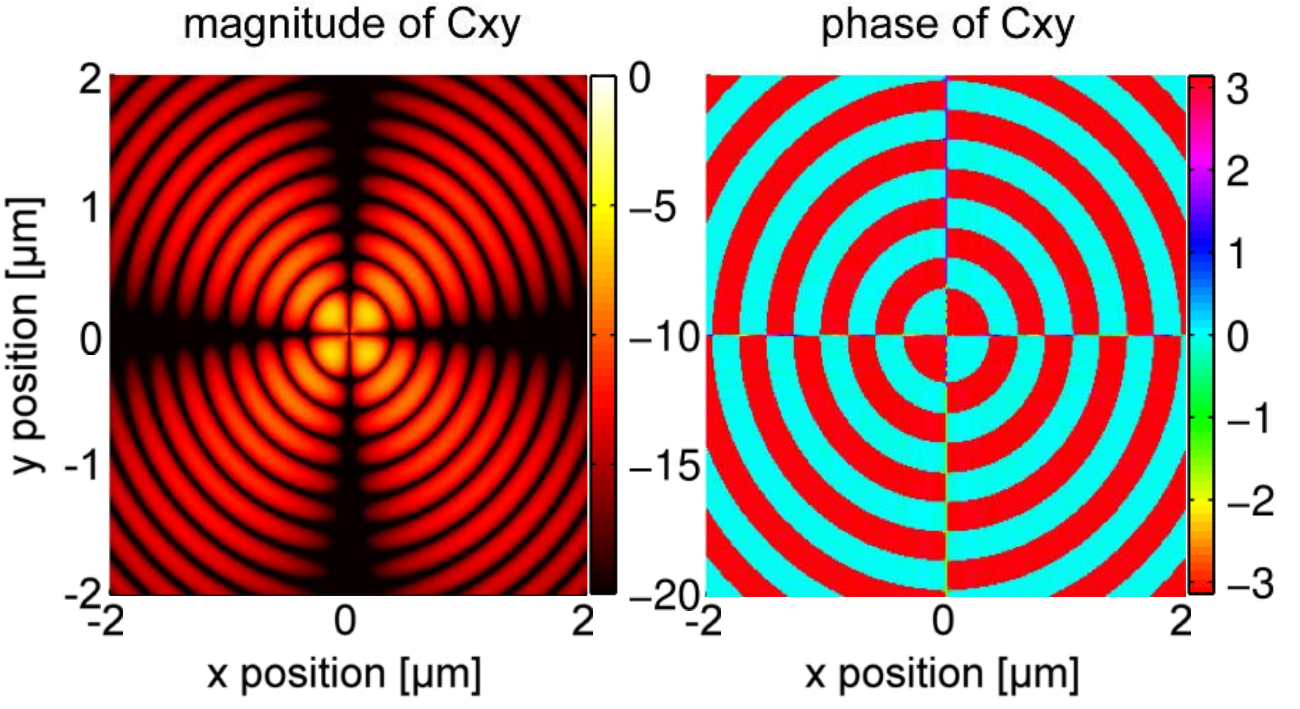} &  &
\includegraphics[width=8.5cm]{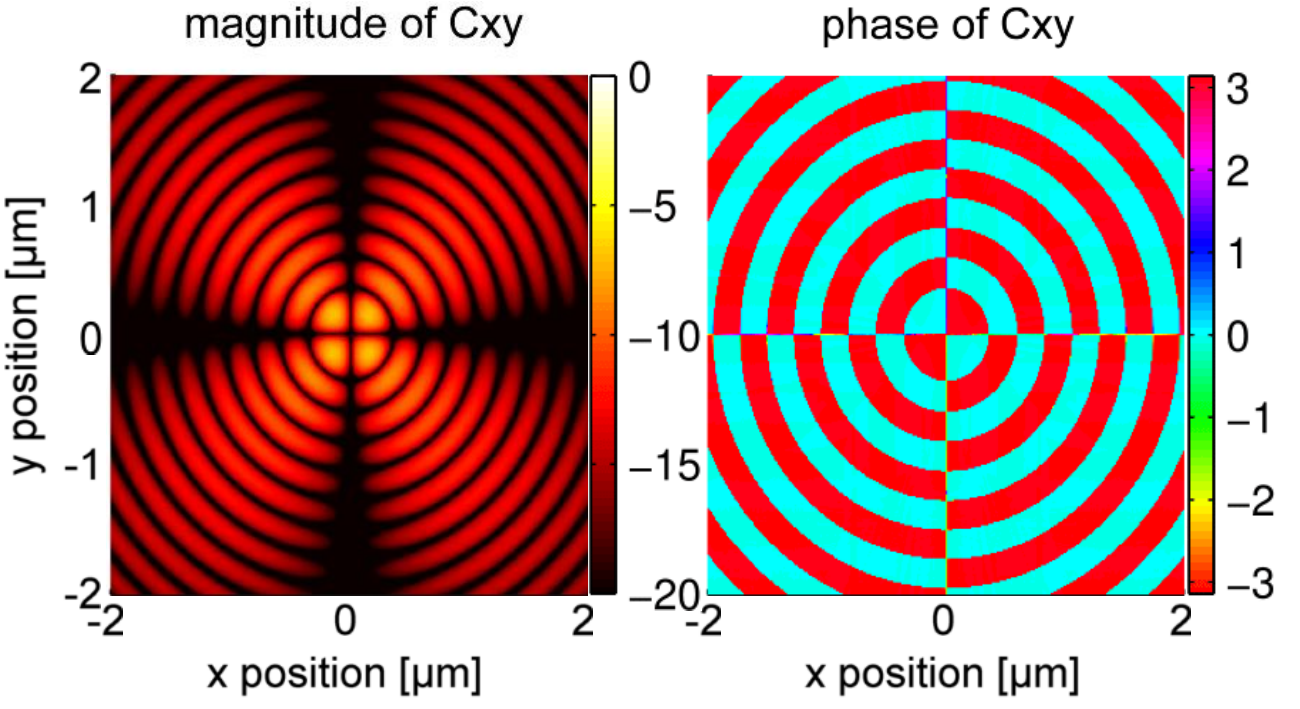} \\
(c) &  & (d) 
\end{tabular}\vspace{-0.3cm}
\caption{$\Cmn(\rpos, \rposp)$ of the silver slab as a function of $\rposp$ for sources located 10 nm (a,c) and 25 nm (b,d) above the surface and $\rpos \cdot \hatx = \rpos \cdot \haty = 0$. Each experiment ((a,c) and (b,d)) has been normalized according to the maximum value of $\Cmn$. The magnitude is provided in dB scale.}
\label{fig:Cmn_slab}
\vspace{0.8cm}
\begin{tabular}{ccc}
\vspace{-0.2cm}
\includegraphics[width=8.5cm]{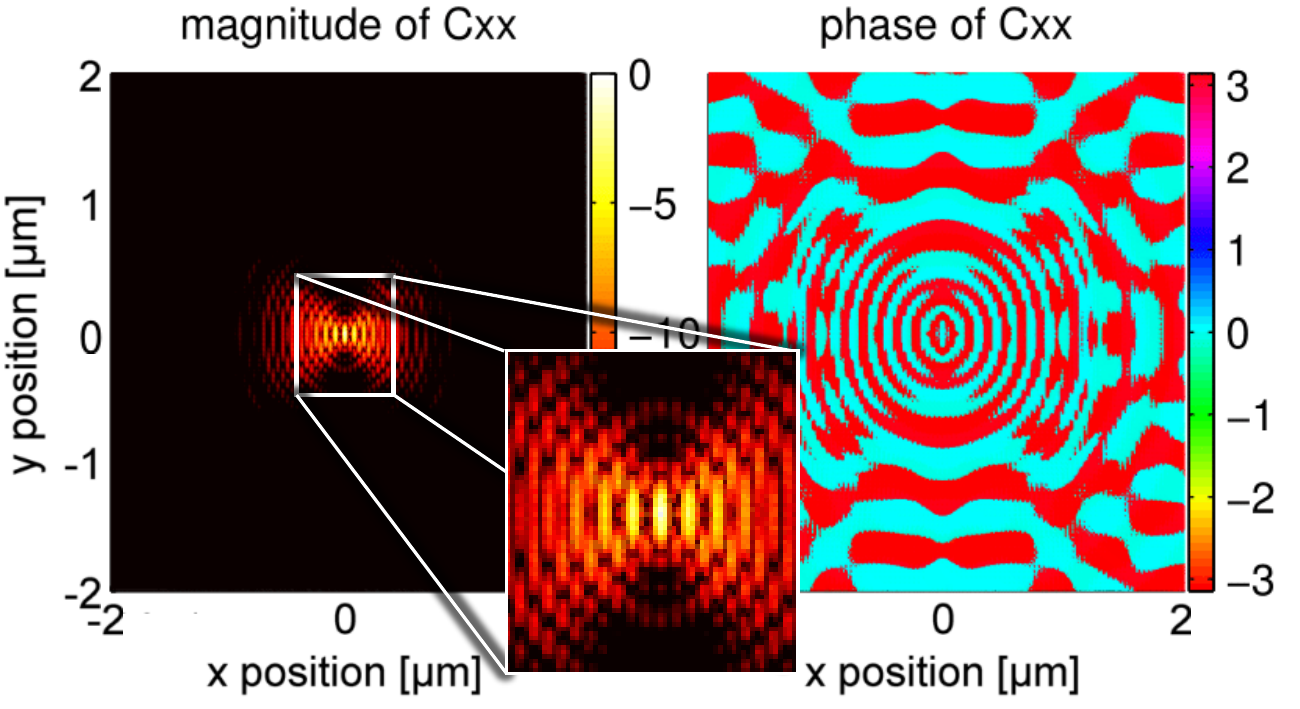} &  &
\includegraphics[width=8.5cm]{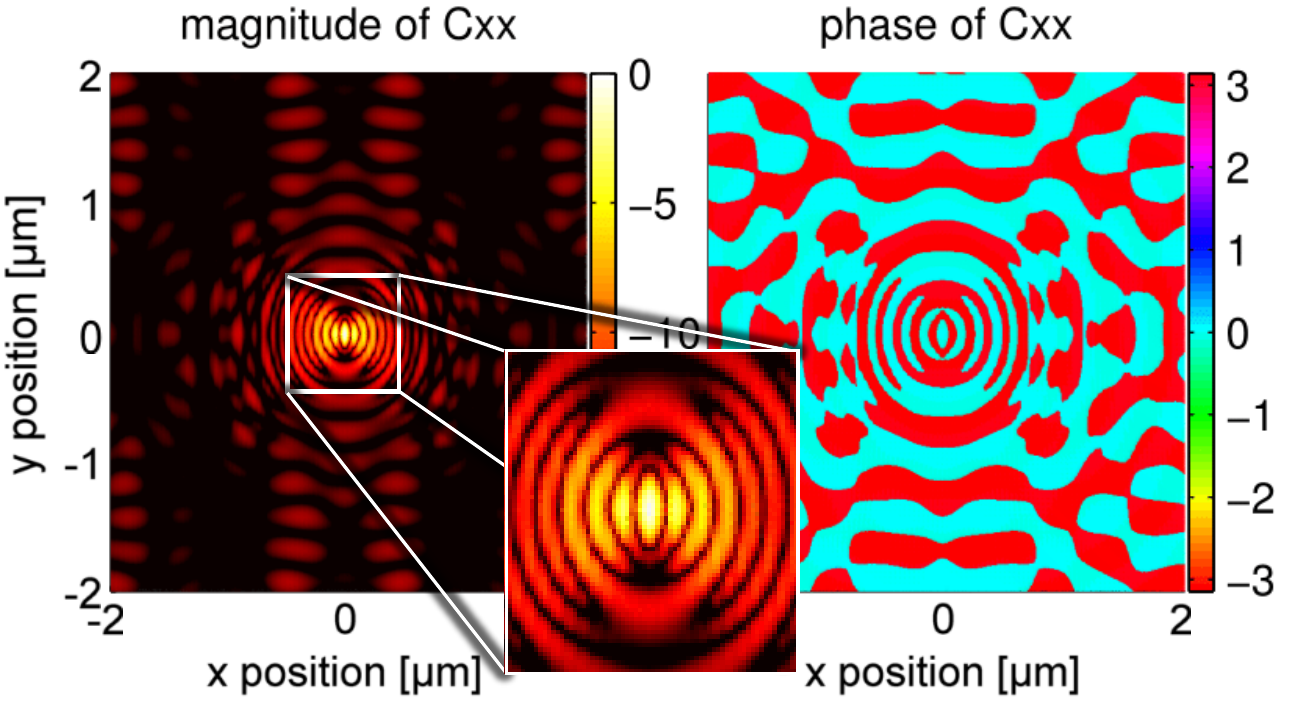} 
\\
\vspace{0.2cm} (a) &  & (b) \\
\vspace{-0.2cm}
\includegraphics[width=8.5cm]{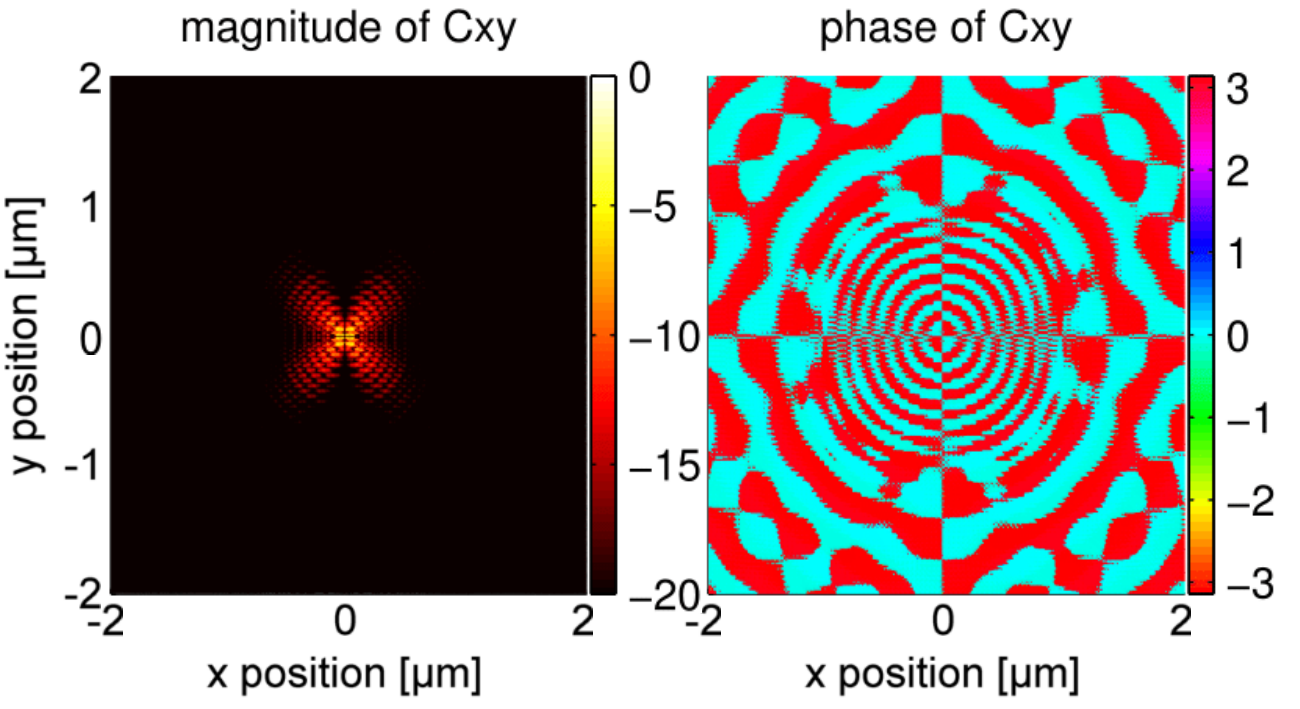} &  &
\includegraphics[width=8.5cm]{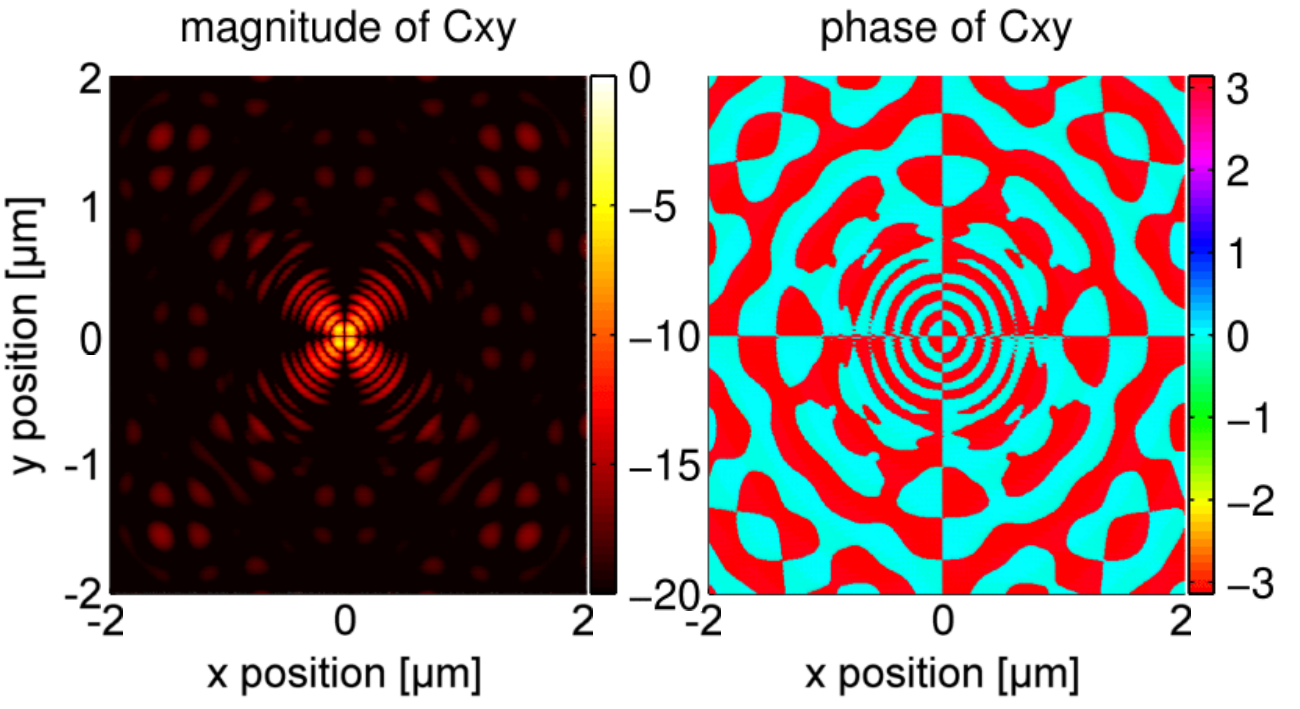} \\
 (c) &  & (d) 
\end{tabular}\vspace{-0.3cm}
\caption{$\Cmn(\rpos, \rposp)$ of the wire medium \cite{Ono2005} as a function of $\rposp$ for sources located 10 nm (a,c) and 25 nm (b,d) above the surface and $\rpos$ shifted 20 nm away from the center of one rod in the $-\hatx$ direction. Each experiment ((a,c) and (b,d)) has been normalized according to the maximum value of $\Cmn$. The magnitude is provided in dB scale.}
\label{fig:Cmn_wire}
\end{figure*}

\begin{figure}[h!]
\center
Path followed \\
\includegraphics[width = 3.5cm]{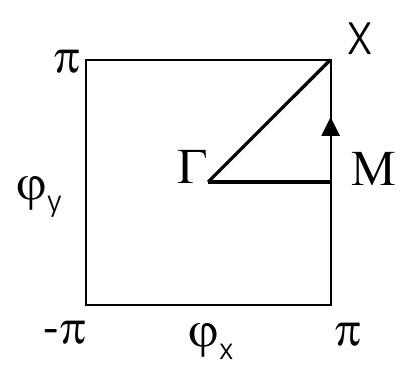}
\includegraphics[width=8cm]{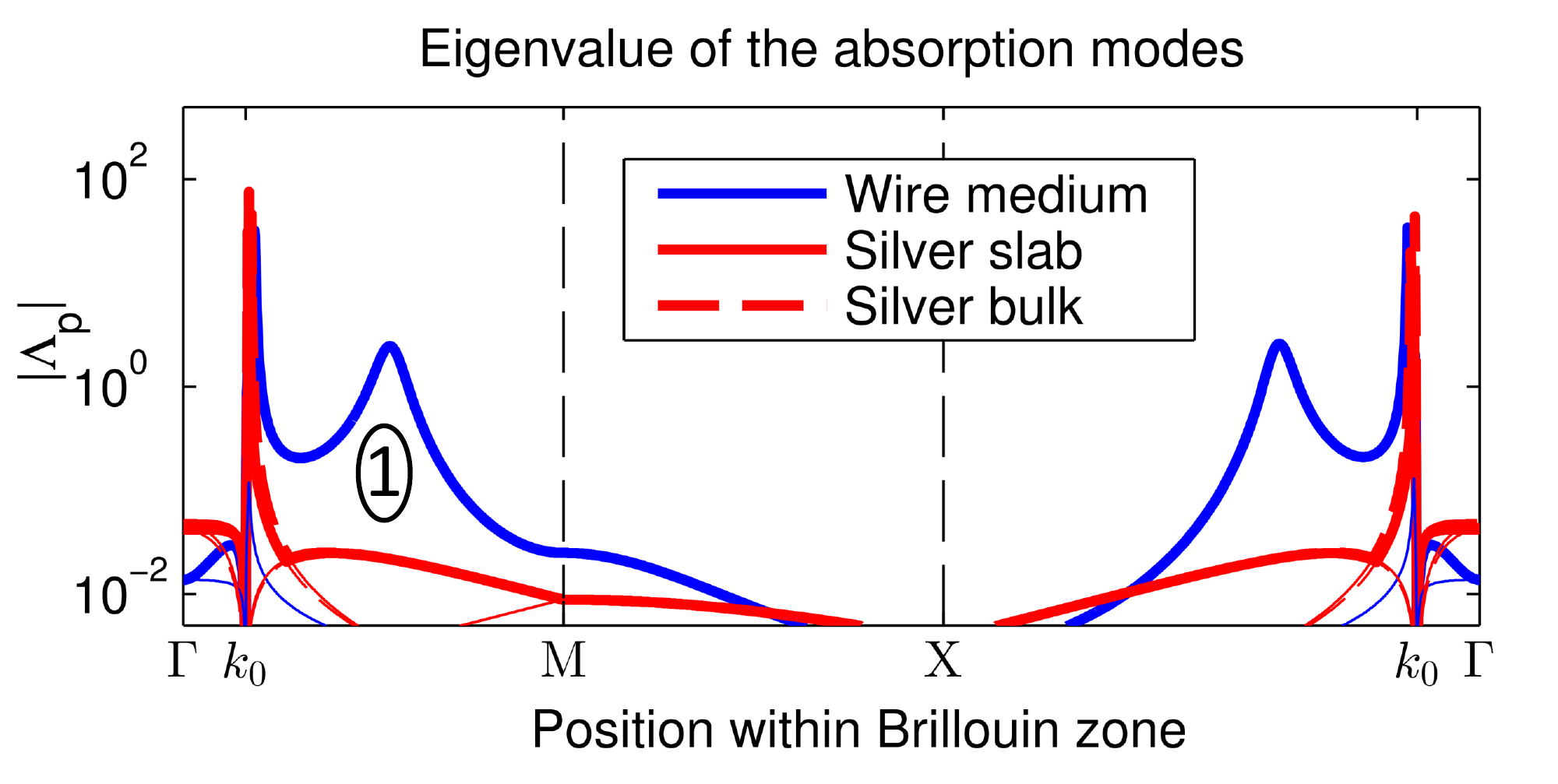}
\includegraphics[width=8cm]{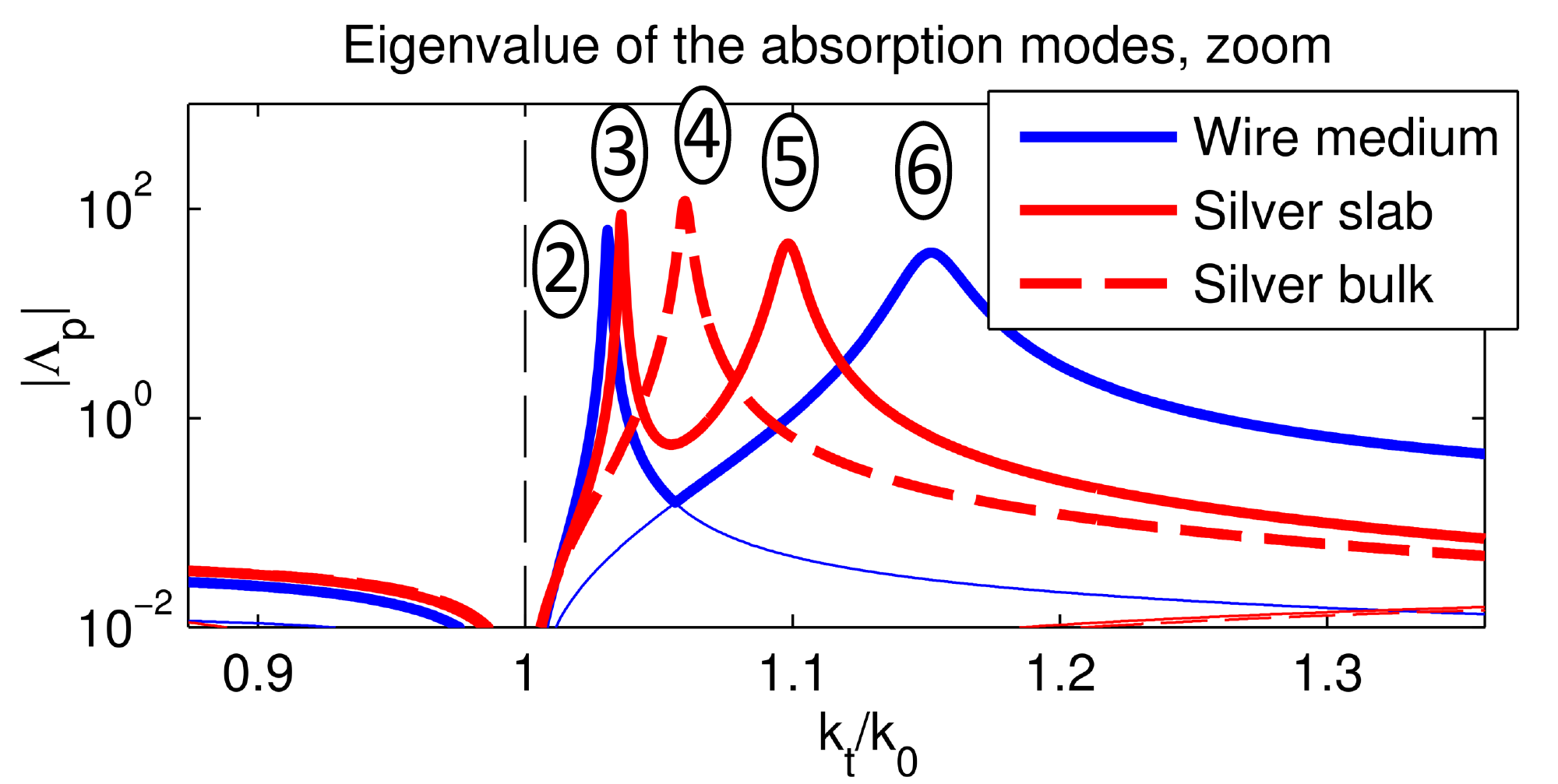}
\caption{Evolution of the amplitude of the absorption modes $\eigP$ within the first Brillouin zone. For better readability, The curve corresponding to the maximum value of $\eigP$ for each phase-shift is thicker than the others.}
\label{fig:path}
\end{figure}

First, we computed the $\Cmn$ function for both absorbers (cf. Section \ref{sec:diff_exc}). The structure is excited using two dipoles located in positions $\rpos_1$ and $\rpos_2$. The associated currents distribution reads $\mathbf{J}_i(\rpos) = \hat{\mathbf{p}}_i \delta(\rpos-\rpos_i)$ with $\rpos_i$ the position of source $i$ and $\hat{\mathbf{p}}_i$ its orientation. The orientation of the sources is either along \xdir or \ydir. The first source is located at $(x,y) = (0,0)$ and oriented along \xdir. The amplitude and phase of the correlation function is displayed as a function of the position and orientation of the second source in Figs. \ref{fig:Cmn_slab} and \ref{fig:Cmn_wire} for sources located 10 and 25 nm above the surface.
We first analyse Fig. \ref{fig:Cmn_slab}.  

In Figures 4(a) and 4(c), we can notice that absorption is mediated by a long-range process based on the plasmonic response of the silver slab. Looking at the phase, the $\Cmn$ function being real is a direct consequence of the translational symmetry of the interface. The observed wavelength of the oscillations in the correlation function is 450 nm, which is slightly shorter than the free-space wavelength of the excitation. Since the shape of this mode can be matched with slightly evanescent incident waves, this absorption mechanism is still limited to the near-field, but not the very near-field, as indicated by the similarity of the patterns for sources located 10 and 25 nm away. 
This process extends over large distances and any surface engineering to improve absorption should target similar length scales.

Now, looking at the results for the wire medium in Fig. \ref{fig:Cmn_wire}, we can first see that the response is much more localized. Indeed, the coupling between two sources vanishes very rapidly. It is due to the fact that the dipoles tend to excite very localized resonant modes along the wires \cite{Ono2005}. 

Interestingly, the microstructure is still clearly visible when sources are located 10 nm away from the surface (see inset in Fig. \ref{fig:Cmn_wire}(a)), while the response is much smoother for sources located 25 nm away from the surface (see inset in Fig. \ref{fig:Cmn_wire}(b)). It is due to the exponential term in (\ref{eq:20-07-01}), which is equal to 0.27 in the first case and reduces to 0.038 in the second case. In other words, the patterning of the surface being deeply subwavelength, variations within one unit-cell are only visible in the very near-field.

Last, we can see some minor ripples appearing in the graphs (b) and (d). These ripples have a longer wavelength than the hot spot in the center of the figure and correspond to additional long range surface waves that are also excited by the sources, but whose contribution to the absorption is less important. 
The geometrical pattern exhibited by this long-range mode is unusual for classical surface modes. It is due to simulation artifacts. Indeed, in order to simulate the presence of two sources in front of a periodic structure the Array Scanning Method has been used, with 250x250 phase shifts \cite{Munk1979}. It means that the surface was not excited by dipoles, but rather by arrays of dipoles whose periodicity is 10 $\mu$m. Since the second mode is a long-range mode, as will be shown later in Table \ref{tab:peak_param}, interference due to the spurious replicas of the sources are appearing, leading to the strange pattern that can be seen in Figs. \ref{fig:Cmn_wire}(b) and \ref{fig:Cmn_wire}(d). However, increasing the number of phase-shifts considered in the ASM, which corresponds to an increase of the periodicity of the sources, or adding losses in the air in order to reduce the propagation length of the mode, the typical length of the oscillation due to this long-range surface mode is preserved while the geometrical pattern exhibited converges to a more classical shape of long-range surface mode.

In order to further analyse these modes, the eigenvalues $\eigP$ of (\ref{eq:14-04-05}) have been plotted for various phase shifts between consecutive unit cells. They have been computed for both the previous two absorbers (wire medium and finite silver slab), as well as for a semi-infinite silver slab (single air-silver interface). The latter is used as a reference to be compared with the analytical results of \cite{SPP} (see below). The phase shifts have been chosen along a path in the first Brillouin zone of the lattice, which can be seen in Fig. \ref{fig:path}, together with the results. The peaks that can be seen correspond to eigenmodes of the structures with associated transverse wave vector and attenuation constant \cite{Zheng2013}. These modes correspond to poles in the complex plane for the fields excitation. Since the total power absorbed is a quadratic quantity of the fields, the peaks correspond to Lorentzian curves whose position and thickness provides information about the complex wavenumber of the modes. 

Firstly, for the wire medium, we can see a large but moderately high peak for transverse wave-vectors that are around $3k_0$ (peak 1). This peak corresponds to the main near-field absorption mechanism of the wire medium. Its large width is the signature of a localized mode (high attenuation constant), while its position correspond to the wavelength of the oscillations that can be seen in Fig. \ref{fig:Cmn_wire}. Zooming around the visibility limits, we can see one high and narrow peak for the silver-air interface (peak 4), and two pairs of narrow peaks for the wire medium (peaks 2 and 6) and the silver slab (3 and 5). The former corresponds to the surface plasmon propagating along a silver-air interface, while the latters correspond to the long-range coupling processes that can be seen in Fig. \ref{fig:Cmn_wire}(b) and (d) for the wire medium slab and Fig. \ref{fig:Cmn_slab} for the silver slab. 

The presence of two surface modes whose transverse wave-number is quite close, with one mode that is more lossy than the other, is typical for thin films supporting surface waves \cite{Sarid1981, Inagaki1985}. Each interface between the structure and the air is supporting a surface mode with an associated penetration depth of the field that is bigger than the actual thickness of the layer. Since the fields of the two surface modes are overlapping, an hybridization of the modes occurs. On one hand, the symmetric coupling of the two surface waves corresponds to a mode for which the field inside the slab is increased (constructive interferences). Therefore, the corresponding modes will be characterized by a higher transverse wavenumber and higher losses (peaks 5 and 6). On the other hand, the antisymmetric coupling of the surface waves corresponds to a mode for which the fields inside the slab are low (destructive interferences), resulting in a lower transverse wavenumber and lower losses (peaks 2 and 3). 

A quantitative analysis of these modes can be performed considering that each peak correspond to a Lorentzian curve. By determining the position and width of the Lorentzian curve, the wave number associated to the modes can be found and are displayed in Table \ref{tab:peak_param}. Comparing the wavenumber obtained in that way with the theoretical wave number of the surface plasmon of the silver slab \cite{SPP}, it can be seen that the agreement is very good ($1.06-0.0023j$ vs. $1.06-0.0022j$).

\begin{table}
\center
\begin{tabular}{|c|c|}
  \hline
  mode index & normalized wave number\\
  \hline
  1 & $3.315 - 0.269j$ \\ 
  2 & $1.031 - 8.66\times10^{-4}j$ \\
  3 & $1.036 - 7.51\times10^{-4}j$ \\
  4 & $1.060 - 0.0023j$ \\
  5 & $1.098 - 0.0050j$\\
  6 & $1.152 - 0.014j$ \\
  \hline
\end{tabular}
\caption{Normalized wave number of the surface modes corresponding to the peaks in Fig. \ref{fig:path}.}
\label{tab:peak_param}
\end{table}

\section{Design of an absorber with a scattering interface layer}
Knowing the response of the silver slab and the wire medium to external excitations, one can try to optimize the absorption of the structure for propagative plane waves, i.e. with a transverse wavenumber smaller than $k_0$. To do so, a scattering layer made of a periodic grating is added on top of the absorber in order to couple propagative plane waves to the evanescent surface waves. 

First, one has to chose between the surface slab and the wire medium. Roughly speaking, the grating on top the surface will scatter the incident fields into each of the available modes, both upward (in the air) and downward (in the silver slab or the wire medium). If the periodicity of the grating is small enough (smaller than $\lambda/2$), only two of these modes, corresponding to the two possible polarizations, propagate in the air, providing only two degrees of freedom for the system to evacuate the incident energy without dissipating it. Therefore, a higher number of available lossy modes in the wire medium or the silver slab should provide a higher absorption of the incident fields \cite{Narimanov2010, Guclu2012}.

To highlight the difference in the number of modes in the air and in the wire medium, we consider artificially big ``unit" cells, which we call super-cells, each of them containing an array of 6 by 6 rods. The size of the super-cell is 240x240 nm$^2$, which is close to $\lambda/2 \times \lambda/2$. In this way, the ratio between the number of modes in the air (maximum 2) and the number of modes in the wire medium appears more clearly.
Artificially reducing the symmetries of the structure in this way adds complexity in the solution, so that some Floquet modes that were decoupled considering all the symmetries of the structure cannot be \textit{a-priori} decoupled anymore in the super-cell based structure (see \eqref{eq:19-07-01} with $a$ and $b$ that match the size of the super-cell). It results in a spectral folding of the absorption modes since a whole family of phase shifts $\bm{\varphi}^{UC}_{mn}$ between consecutive unit cells corresponds to the same phase shift $\bm{\varphi}^{SC}$ between super-cells:
\begin{equation}
\bm{\varphi}^{UC}_{mn} = \dfrac{\bm{\varphi}^{SC}}{6} + \Big( m\dfrac{\pi}{3}; n \dfrac{\pi}{3} \Big), ~~~~ m, n \in \{0,1,2,3,4,5\}.
\end{equation}

In order to quantitatively compare the potential absorption that the wire medium and the silver slab can provide, the eigenvalues of the 15 highest absorbing modes within 240x240 nm$^2$ super-cells are plotted for the wire medium and the silver slab for the $\Gamma-X-M-\Gamma$ path within the first Brillouin zone (see Fig. \ref{fig:path}). 

\begin{figure}[h!]
\center
\includegraphics[width = 9cm]{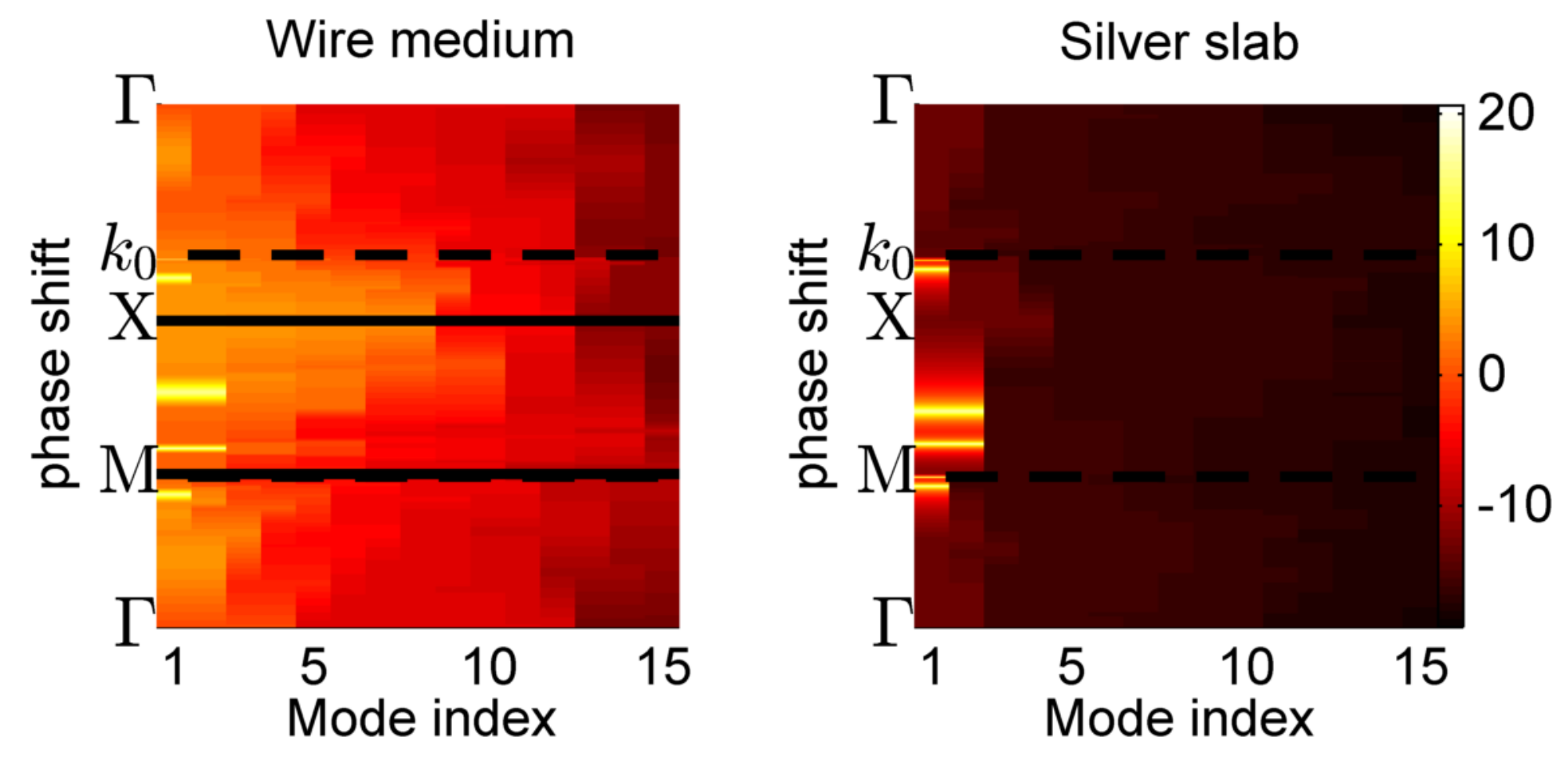}
\caption{Magnitude of the eigenvalue $\eigP$ of the first 15 eigenmodes for the wire medium and the silver slab. The magnitude is in dB scale.}
\label{fig:mode_per_uc}
\end{figure}

It can be clearly seen that the silver slab has a great potential to selectively absorb the fields coming from particular incidence angles, while the wire medium is potentially much more efficient for non-selective absorption. For that reason, the latter will be used as a substrate for the improved absorber.

\begin{figure}[h!]
\center
\includegraphics[width = 6cm]{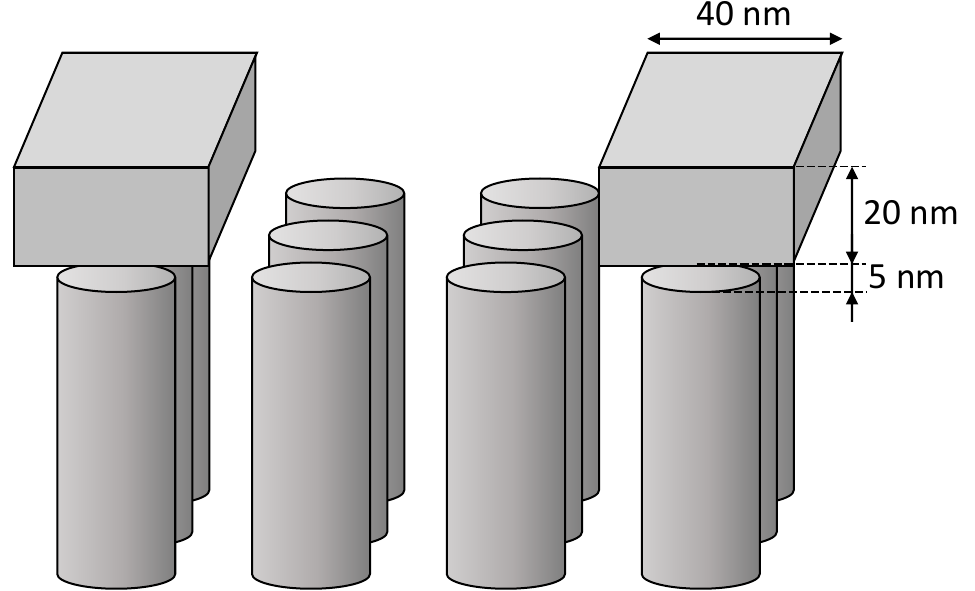}
\caption{Geometry of the improved absorber.}
\label{fig:geom_scatt}
\end{figure}

Next comes the question of the periodicity of the grating. The grating will correspond to silver strips aligned with the \ydir direction and deposited 5 nm above the top of the wires. The trenches are 40 nm wide and 20 nm thick. In order to simplify the problem, the trench is replicated every $n$ wires, $n$ being an integer. The geometry is drawn in Fig. \ref{fig:geom_scatt} for $n=3$. 

\begin{figure}[h!]
\center
\includegraphics[width = 5cm]{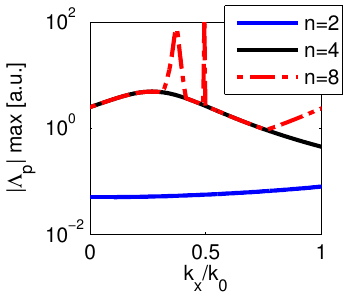}
\caption{Maximum absorption mode to which propagating plane wave can couple if the period of the scatterer is $n$ times that of the underlying absorber in the \xdir direction.}
\label{fig:spectral_folding}
\end{figure}

Increasing the periodicity of the unit cell using a grating will introduce spectral folding, so that absorption modes of the wire medium that cannot couple with propagating plane waves can be folded back into the visible zone and therefore be coupled to the incident waves through the periodic grating, as proposed in \cite{Argyropoulos2013, Meng2014, Zhao2014}. In figure \ref{fig:spectral_folding}, the maximum value $\eigP$ with which incident waves with transverse wave vector $\kt = (k_x, 0)$ can couple for a periodicity of the grating $n=2, 4$ and $8$ has been drawn. It can be seen that for $n=4$, the localized mode (peak 1) has been folded back into the visible zone, while for $n=8$ sharp peaks corresponding to the spectral folding of the long-range surface modes of the wire medium (peaks 2 and 4 in Fig. \ref{fig:path}(c)) are also visible.

\begin{figure}[h!]
\center
\begin{tabular}{cc}
\includegraphics[width = 4cm]{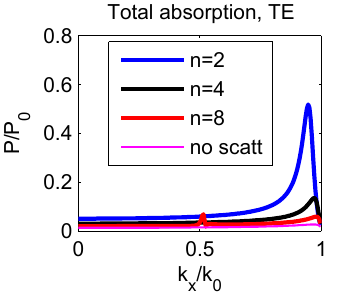}
&
\includegraphics[width = 4cm]{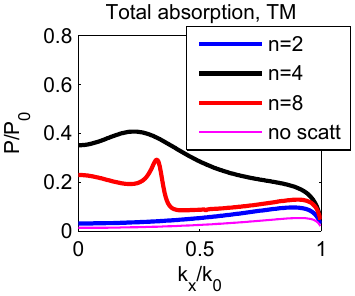}
\\
(a) & (b) 
\end{tabular}
\caption{Normalized absorption of incident TE (a) and TM (b) plane waves by the improved absorber for varying incidence angles and periods of the grating.}
\label{fig:abs_scatt}
\end{figure}

A full-wave study of the plane wave absorption with the periodic grating of periodicity $n=2, 4$ and $8$ has also been performed and its results are displayed in Fig \ref{fig:abs_scatt}. First of all, there is an unexpected increase in the absorption for TE waves near grazing incidence for $n=2$. Since half of the power is absorbed within the grating, it corresponds to some hybrid absorption mode that appears due to the coupling between the wire medium and the grating. This behaviour cannot be predicted from the sole study of the wire medium. A small glitch around $k_x \simeq 0.5 ~ k_0$ is also noticeable for $n=8$. It corresponds to the coupling between the grating and one surface mode of the wire medium (here, mode 2 in Fig. \ref{fig:path}). This effect is much more visible for TM incident waves: looking at the absorption of the TM waves, a strong similarity between Fig. \ref{fig:spectral_folding} and Fig. \ref{fig:abs_scatt}(b) is noticeable. Indeed, for TM waves, most of the absorption is taking place within the wire medium. It means that the grating is scattering the incident fields into the absorption modes of the wire medium. The latter dissipates a non-negligible part of the incident energy. It can be noticed that the dissipation in the case $n=8$ is lower than in the case $n=4$. It is probably due to the fact that the scattering cross section of the grating decreases as the trenches are placed farther away. Increasing the covering by the scatterer without modifying the periodicity could potentially solve the issue.

\section{Conclusion}
The complete characterization of an absorbing structure can be a difficult task if not properly handled due to the fact that the power is not a linear function of the incident fields. However, it can be shown that any structure absorbs energy through a discrete number of independent modes. In this paper, we first showed that the plane wave absorption spectrum of the structure does not provide a sufficiently rigorous analysis of the absorption capabilities that structure. This fact was illustrated through a practical example consisting of an array of resonant golden patches on top of a grounded dielectric, whose period is greater than half a wavelength. Then we used Energy Absorption Interferometry to fully characterize two different plasmonic absorbers: a silver slab and an array of parallel silver rods. 

For the silver slab, the absorption is predominantly mediated by long-range surface modes that correspond to hybridized surface plasmons supported by the air-silver interfaces. However, these modes being long-range, their exploitation must be thought of on a large scale.
For the wire medium, different competing processes were highlighted for near field absorption, two long-range surface modes and one resonant localized mode. By adding some scatterers on top of the wire medium, incident plane waves can be coupled to some of these modes, leading to an absorption increased by more than one order of magnitude. The coupling to one or the other of the modes can be engineered by properly changing the period of the scatterer. It was shown that the processes leading to long-range correlation in the absorption can lead to spatially selective absorbers, while the localized resonant modes lead to a non-selective increase in the absorption. 

\section*{Acknowledgments}
Computational resources have been provided by the Consortium des Equipements de Calcul Intensif (C\'{E}CI), funded by the Fonds de la Recherche Scientifique de Belgique (F.R.S.-FNRS) under Grant No. 2.5020.11.

\end{document}